\documentclass[aip,jcp,reprint]{revtex4-1}

\usepackage{graphicx}
\usepackage{amsmath}
\usepackage{amssymb}
\usepackage{multirow}

\newcommand{\Angstrom}{\ensuremath{\mathring{\textnormal{A}}}}
 \newcommand{\sub}[1]{\ensuremath{_{\textrm{#1}}}} \newcommand{\super}[1]{\ensuremath{^{\textrm{#1}}}} \newcommand{\sci}[2]{\ensuremath{#1 \times 10^{#2}}}

\begin{document}

\title{Weighted-density functionals for cavity formation and dispersion energies in continuum solvation models}

\author{Ravishankar Sundararaman, Deniz Gunceler, and T A Arias}
\affiliation{Cornell University Department of Physics, Ithaca, NY 14853, USA}
\date{\today}

\begin{abstract}
Continuum solvation models enable efficient first principles calculations of
chemical reactions in solution, but require extensive parametrization and fitting
for each solvent and class of solute systems. Here, we examine the assumptions
of continuum solvation models in detail and replace empirical terms with
physical models in order to construct a minimally-empirical solvation model.
Specifically, we derive solvent radii from the nonlocal dielectric response
of the solvent from \emph{ab initio} calculations, construct a closed-form and
parameter-free weighted-density approximation for the free energy of the cavity formation,
and employ a pair-potential approximation for the dispersion energy.
We show that the resulting model with a single solvent-independent parameter:
the electron density threshold ($n_c$), and a single solvent-dependent parameter:
the dispersion scale factor ($s_6$), reproduces solvation energies of organic molecules
in water, chloroform and carbon tetrachloride with RMS errors of 1.1, 0.6 and 0.5 kcal/mol
respectively. We additionally show that fitting the solvent-dependent $s_6$ parameter to
the solvation energy of a single non-polar molecule does not substantially increase these errors.
Parametrization of this model for other solvents, therefore, requires minimal effort
and is possible without extensive databases of experimental solvation free energies.
\end{abstract}

\maketitle

\makeatletter{}Most chemical reactions of technological interest and most biological processes occur
in a liquid environment, and the solvent plays a critical role in determining reaction pathways.
Direct treatment of solvent molecules in first principles studies of these systems
is cumbersome due to the large number of electrons that must be included and 
the number of geometries necessary to adequately sample the phase space of the liquid.
Continuum solvation models that approximate the solvent effects by the response of a
continuum dielectric along with empirical corrections enable efficient calculations
of systems in solution.

Several variants of continuum solvation models, notably the
polarizable continuum models\cite{PCM94,PCM97,PCM-Review} (PCMs)
and the `SMx' series of solvation models,\cite{PCM-SM1,PCM-SM8,PCM-SMD}
are widely available in quantum chemistry software, and have been successfully
applied to understand molecular reaction mechanisms and improve homogeneous catalysts.
In order to enable similar studies of heterogeneous catalysis,
several solvation models have been recently developed for periodic systems
(particularly in software employing plane-wave basis sets) including the self-consistent
continuum solvation (SCCS) models\cite{PCM-SCCS,PCM-SCCS-charged} and the simplified
models\cite{JDFT,PCM-Kendra,NonlinearPCM} derived from joint density-functional theory (JDFT).

All these solvation models place the solute system (typically treated with
electronic density-functional theory) in a dielectric cavity in order to
describe the dominant electrostatic interactions between solute and solvent,
and then include corrections for other contributions such as the free energy
of forming a cavity in the solvent and dispersion interactions.
The models differ in the determination of the cavity shape and size for a given
solute system and in the approximations to the cavity formation and dispersion energies.

Traditional chemistry solvation models adopt atom-based parametrizations.
The PCM approach\cite{PCM-Review} defines a `solvent accessible surface' (SAS)
as the boundary of the region of space accessible to the centers of the solvent molecules,
which is carved out by spheres sized by radii of the solute atoms and of the solvent molecule.
Additionally, the smaller `solvent excluded surface' (SES) bounds the region of space
which overlaps with any solvent sphere with center placed on or outside the SAS.
The dielectric cavity is typically obtained by expanding the SES by a solvent-dependent radius
and the cavitation and dispersion energies are computed using the SAS, either using
empirical surface tensions or employing combinations of scaled particle theory
and pair-potential dispersion corrections. (See Ref.~\citenum{PCM-Review} for details.)
These models require a number of atom-dependent parameters such as atomic radii or
effective surface tensions, which limits their applicability to the class
of solutes that they are fit to: typically organic molecules and ions.
The atom-based parametrization is, however, mostly transferable between solvents
and requires only a few solvent-dependent parameters. (See Ref.~\citenum{PCM-SMD}, for example.)

On the other hand, the solvation models used in conjunction with plane-wave basis sets
predominantly employ an electron density based parametrization. The `cavity' is actually
a smooth variation of the dielectric constant as a function of the electron density:
from 1 in the high electron-density `solute' region of space to the bulk solvent
dielectric constant $\epsilon_b$ in the low electron-density `solvent' region of space.
These models approximate cavity formation and dispersion energies empirically with a term
that depends linearly on the surface area, and optionally also the volume, of the cavity.
The electron-density based models require very few parameters: three to four parameters
for the SCCS models\cite{PCM-SCCS} and two parameters for the simplified JDFT models.\cite{NonlinearPCM}
However, all these parameters inherently include information about the solvent:
the parameter(s) that control the variation of the dielectric constant
encapsulate the size of the solvent molecule, and the remaining parameters
capture cavity formation and dispersion energies, which respectively
depend on the equation of state and polarizability of the solvent
(in addition to microscopic properties of the solvent molecule).
Therefore, these parameters need to be fit separately for each solvent of interest,
using experimental solvation free energies determined from solubility measurements.
We have recently shown\cite{PCM-correlateTau} that it is possible to mitigate this issue
by correlating the parameters used in the solvation model to bulk properties of the solvent.
The use of an empirical surface tension, however, still misses the microscopic
size and shape dependence of the cavity formation and dispersion energy.
Further, for non-polar solvents dominated by dispersion interactions,
this effective tension is negative and presents numerical instabilities
in the self-consistent field approach (electrons leak into the fluid).

Here, we combine the best features of the atom-based and density-based
parametrizations to obtain a continuum solvation model with one universal
(solvent-independent) fit parameter and one fit parameter per solvent,
and which includes a microscopically-accurate model
for the cavity formation and dispersion energies.
Section~\ref{sec:ElectrostaticRadius} analyzes the nonlocality in the solvent response
and presents an \emph{ansatz} to estimate the `electrostatic radius', the spacing between the
dielectric cavity and the solvent-center cavity (SAS), without any fit parameters.
Unlike the atom-centered sphere approaches, it is not trivial to incorporate
this electrostatic radius in the smooth cavities of the electron-density based approaches.
Section~\ref{sec:CavityExpansion} develops a technique to `expand' electron densities
by a given radius, which enables the determination of the dielectric cavity as well as the SAS
using a single function of the electron density. This function contains the single universal
fit parameter $n_c$, a characteristic electron density at which the cavity transitions.

Sections~\ref{sec:CavitationModel} develops a closed-form weighted-density approximation
for the free energy of cavity formation which depends only on experimentally measurable
bulk properties of the solvent (no fit parameters), and shows that it
accurately reproduces results obtained from classical density-functional theory.
Section~\ref{sec:DispersionModel} adapts the pair-potential dispersion corrections\cite{Dispersion-Grimme}
used in electronic density-functional theory to describe the dispersion interactions
between the solute and solvent. This term introduces the single solvent-dependent
parameter $s_6$, an empirical scale factor for the dispersion corrections.
Finally, Section~\ref{sec:SolvationEnergies} presents the fit parameters
and solvation energy results of the resulting continuum solvation model
for water, chloroform and carbon tetrachloride, and demonstrates the plausibility
of almost solvent-independent parametrization of simplified solvation models.
 
\makeatletter{}\section{Electrostatic radii of solvents} \label{sec:ElectrostaticRadius}

The fundamental need for empirical parameters in polarizable continuum models
arises from the locality assumption: the nonlocal response of the solvent
is replaced by that of a continuum dielectric. This assumption is compensated
for by choosing the boundary of the dielectric appropriately: the optimum
boundary lies between the SAS (solvent centers) and SES (solvent edges)
and is selected by fitting some cavity size parameter, critical electron density $n_c$
or radius scale factor depending on the approach, to solvation energies.
Here, we analyze the nonlocal response of the solvent, obtained from electronic
density-functional calculations of one solvent molecule, to determine the distance
of this optimum dielectric boundary from the solvent-center surface (SAS).

We start by computing the charge density of a single solvent molecule $\rho\sub{mol}(\vec{r})$
using electronic density-functional theory, and expanding its electronic and vibrational susceptibility
obtained from density-functional perturbation theory in an eigen-basis
\begin{equation}
\chi\sub{mol}(\vec{r},\vec{r}') = -\sum_i X_i \rho_i(\vec{r}) \rho_i(\vec{r}')
\end{equation}
with normal modes of strength $X_i$ with characteristic charge density $\rho_i(\vec{r})$.

Next, consider a single solvent molecule with its center pinned at the origin
that is free to rotate and in thermal equilibrium at temperature $T$.
In the absence of any external fields, this molecule adopts all orientations
$\omega \in \textrm{SO}(3)$ with equal probability $p_\omega = 1$
(normalized so that $\int \frac{d\omega}{8\pi^2} p_\omega = 1$).
With a perturbing field, the orientation density is altered to first order
in the field and in each orientation, the molecule is polarized by the
field, again to first order. By collecting the total induced charge at first order,
we can show that the net susceptibility of the free rotor at $T$ is
\begin{flalign}
\chi_T(\vec{r},\vec{r}') &= \int\frac{d\omega}{8\pi^2} \left[ 
	\frac{-1}{T} \rho\sub{mol}(\omega\circ\vec{r}) \rho\sub{mol}(\omega\circ\vec{r}) 
	\right. \nonumber\\ &\qquad\qquad\left.
	-\sum_i X_i \rho_i(\omega\circ\vec{r}) \rho_i(\omega\circ\vec{r}')
\right] \nonumber\\
	&= - \int\frac{d\omega}{8\pi^2} \sum_{i=0} X_i \rho_i(\omega\circ\vec{r}) \rho_i(\omega\circ\vec{r}'),
\end{flalign}
where $\omega\circ\vec{r}$ denotes the result of rotating $\vec{r}$ by $\omega \in \textrm{SO}(3)$.
The second line simplifies the notation by extending the sum over polarizability modes
to include rotation as mode 0 with strength $X_0 = 1/T$ and characteristic
charge density $\rho_0(\vec{r}) = \rho\sub{mol}(\vec{r})$.

Finally, in order to estimate the extent of the nonlocality of the response,
we adopt a simple model of the solvent consisting of a fixed distribution
of free-rotor solvent molecules at temperature $T$.
The net susceptibility of a semi-infinite slab ($z>0$) of such a solvent 
with bulk molecular density $N\sub{bulk}$ is therefore
\begin{multline}
\chi(\vec{r},\vec{r}') = - \int d\vec{R} N\sub{bulk}\theta(\vec{R}\cdot\hat{z}) \int\frac{d\omega}{8\pi^2} \\
	\times \sum_{i=0} X_i \rho_i(\omega\circ\vec{r}-\vec{R}) \rho_i(\omega\circ\vec{r}'-\vec{R}). \label{eqn:ElRadiusChi}
\end{multline}
Approximating the interaction between molecules at the mean-field level,
the bound charge density in the solvent is then $\rho\sub{bound} = \hat{\chi} \phi\sub{tot}
\equiv \int d\vec{r}'\chi(\vec{r},\vec{r}') \phi\sub{tot}(\vec{r}')$,
where $\phi\sub{tot}$ is the total electrostatic potential.
For an applied external potential $\phi\sub{ext}$, the total potential
then satisfies the self consistency relation
\begin{flalign}
\phi\sub{tot} &= \phi\sub{ext} + \hat{K}\rho\sub{bound} = \phi\sub{ext} + \hat{K}\hat{\chi}\phi\sub{tot} \nonumber\\
\Rightarrow\quad (1 - \hat{K}\hat{\chi}) \phi\sub{tot} &= \phi\sub{ext},
\label{eqn:ElRadiusSelfCons}
\end{flalign}
where $\hat{K}$ is the Coulomb operator.

\begin{figure}
\includegraphics[width=\columnwidth]{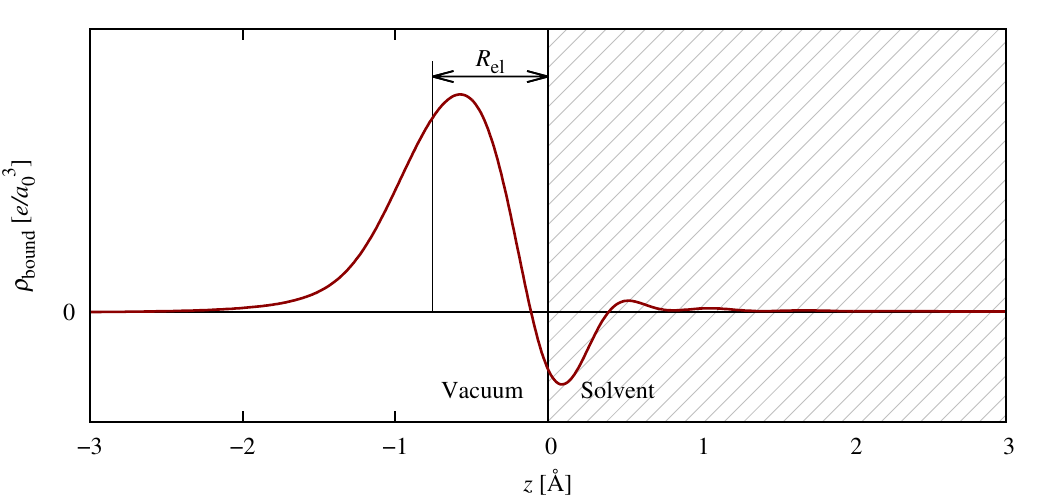}
\caption[Determination of electrostatic radius]
{Determination of electrostatic radius, shown here for water,
as the first moment of the bound charge density at the interface of the solvent slab
with model nonlocal response given by (\ref{eqn:ElRadiusChi}),
under a uniform externally applied field.\label{fig:elRadius}}
\end{figure}

To this model solvent slab, we apply a uniform external field normal to the slab
with $\phi\sub{ext}(\vec{r}) = -Dz$, and numerically solve the one-dimensional integral equation
(\ref{eqn:ElRadiusSelfCons}) with the non-local $\hat{\chi}$ given by (\ref{eqn:ElRadiusChi})
to obtain the total potential and bound charge density in the solvent.
Figure~\ref{fig:elRadius} shows the resultant bound charge density at the interface for liquid water.
In contrast, the bound charge density in a continuum dielectric would be a $\delta$-function
centered at $z=0$.

The interaction energy of this bound charge with a sheet charge
$\sigma \delta(z+L)$ for some large enough $L$ so that the two charge densities
do not overlap is $U\sub{NL} = \sigma \int dz \rho\sub{bound}(z) (z+L)$.
Similarly, the interaction energy of this sheet charge with a
continuum dielectric bounded at $z=-R\sub{el}$ is $U_\epsilon = \sigma (L-R\sub{el}) \int dz \rho\sub{bound}(z)$.
The magnitude of the bound charge for the nonlocal response and a continuum dielectric
with the same bulk dielectric constant are identical, and therefore the interaction
energies $U_\epsilon = U\sub{NL}$ for all $L$ if
\begin{equation}
R\sub{el} \equiv -\frac{\int dz \rho\sub{bound}(z) z}{\int dz \rho\sub{bound}(z)}.
\end{equation}
This defines the electrostatic radius of a solvent, $R\sub{el}$,
as the distance by which a continuum dielectric boundary should be placed
closer to the source charge compared to the solvent-center surface
in order to match the energetics of the nonlocal response in a planar geometry.
At lowest order, this optimum distance is unaffected upon moving from a planar interface
to the cavity geometry around an electronic system, and we set the separation between the
electric response cavity and the SAS to $R\sub{el}$, computed \emph{ab initio} as detailed above.

\begin{table}
\caption{Computed electrostatic radii $R\sub{el}$ and experimental vdW radii $R\sub{vdW}$
from Ref.\citenum{RvdwFluids} for the solvents considered here
\label{tab:Radii}}
\begin{center}\begin{tabular}{ccc}
\hline\hline
Solvent & $R\sub{el}$ [\AA] & $R\sub{vdW}$ [\AA] \\
\hline
H\sub{2}O    & 0.75 & 1.385 \\
CHCl\sub{3}  & 1.17 & 2.53 \\
CCl\sub{4}   & 1.01 & 2.69 \\
\hline\hline
\end{tabular}\end{center}
\end{table}

We calculate the charge density $\rho(\vec{r})$ and the susceptibility $\chi(\vec{r},\vec{r}')$
using the open source plane-wave density-functional theory software JDFTx,\cite{JDFTx}
with the PBE exchange-correlation functional and norm-conserving pseudopotentials
at a kinetic energy cutoff of 30~$E_h$ ($\approx 816$~eV).
Table~\ref{tab:Radii} lists the eletrostatic radii we obtain using the above procedure
for the three solvents we consider here, water, chloroform and carbon tetrachloride. 
\makeatletter{}\section{Cavity expansion} \label{sec:CavityExpansion}

\begin{figure}
\centering{\includegraphics[width=\columnwidth]{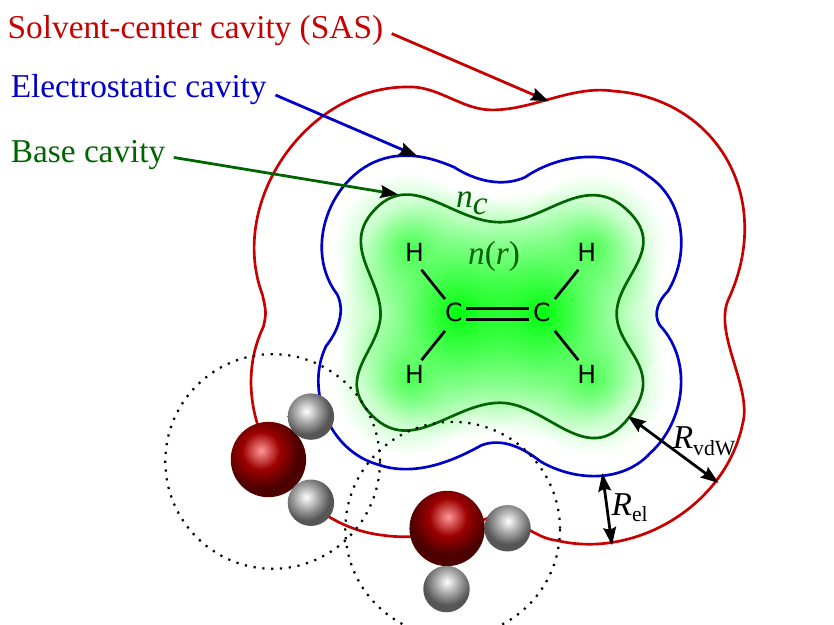}}
\caption[Relation between various cavity surfaces]
{Relation between various cavity surfaces in a PCM description of an ethylene molecule in water.
The solvent accessible surface (SAS) is obtained by expanding the base cavity,
which roughly corresponds to the SES, by the solvent van der Waals radius ($R\sub{vdW}$).
The electric response is smaller than the SAS by the electrostatic radius, $R\sub{el}$,
and is hence obtained by expanding the base cavity by $R\sub{vdW} - R\sub{el}$.
The base cavity is then expected to be a property of the solute alone, and hence solvent independent.
\label{fig:CavityRelations}}
\end{figure}

The electrostatic radius ($R\sub{el}$) defined above links the cavity for electrostatic response
to the surface of solvent centers (SAS). In traditional PCM's, the SAS is determined
from the union of atom-centered spheres of radii equal to the sum of solute atom
and solvent vdW radii. Combining that definition of the SAS with the \emph{ab initio}
computed $R\sub{el}$ would eliminate solvent-dependent scale parameters in the cavity determination.
However, our goal is to take a step forward and avoid atom-dependent parameters, if possible.

Here, we transfer the physical intuition behind the atomic vdW radius approach to
the isodensity approach, where the cavities are determined from the electron density.
vdW radii are defined in terms of the distance of nearest approach of two closed-shell
electronic systems,\cite{RvdwAtoms} and hence are expected to be a reasonable descriptor for
the typical spacing between solute and solvent molecules in the domain of validity
of solvation models; in any case, covalent bonds with the solvent would require
inclusion of the bonded solvent molecules in the quantum-mechanical calculation.

Consequently, we propose the following program (see Figure~\ref{fig:CavityRelations}).
The electron density of the solute is thresholded at a critical density $n_c$
to determine a base cavity, which corresponds roughly to the SES of traditional models.
This cavity is a property of the solute alone, and the resulting $n_c$ can therefore
expected to be independent of the solvent.

The solvent-center cavity used for computing the cavity formation and dispersion energies
is obtained by expanding this base-cavity by the solvent vdW radius, $R\sub{vdW}$,
which can be defined in terms of the equation of state and other thermodynamic properties
of the fluid,\cite{SPT-Review} and has been tabulated for many fluids.\cite{RvdwFluids}
The cavity for electric response is expected to be $R\sub{el}$ smaller than the
solvent-center cavity as discussed previously. We therefore obtain this cavity by expanding
the base cavity by $R\sub{vdW} - R\sub{el}$.

\begin{figure}
\begin{tabular}{cc}
\includegraphics[width=0.48\columnwidth]{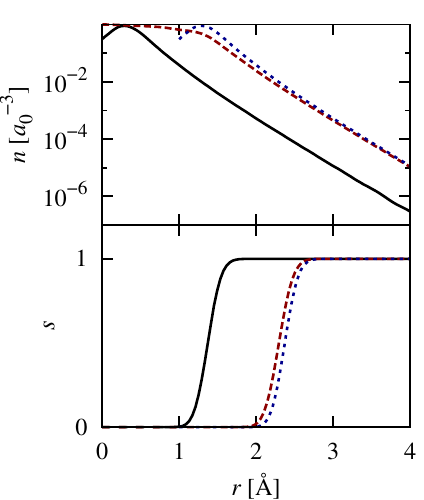}&\includegraphics[width=0.48\columnwidth]{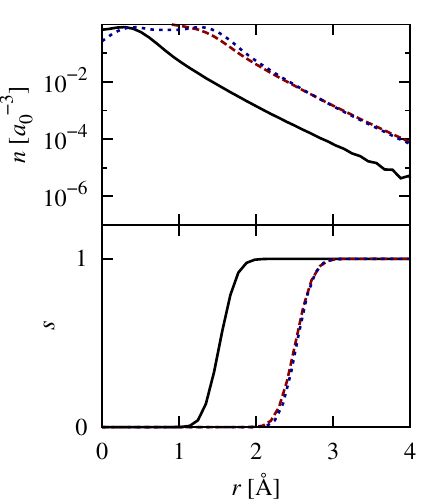} \\
 (a) \includegraphics[width=0.75in]{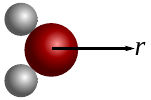} & (b) \includegraphics[width=1.25in]{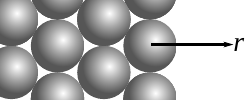}
\end{tabular}
\caption[Accuracy of electron density expansion functional]
{Accuracy of electron density expansion functional for
(a) water with an approximately spherical electron density tail
and (b) Pt(111) surface with an approximately planar electron density tail.
In the upper (lower) panels, the solid black lines show the original electron density (cavity),
the dashed red lines plot the corresponding quantities for the electron density expanded
by $1$~\Angstrom, and the dotted blue lines plot the original quantity shifted outwards by $1$~\Angstrom.
\label{fig:nExpansion}}
\end{figure}

At this stage, all cavities required for the model can be determined from
a single critical electron density $n_c$,
along with thermodynamic and \emph{ab initio} computed properties of the solvent.
In the rest of this section, we develop a practical method to construct the
expanded cavities in a plane-wave basis calculation.

Only the exponential tail regions of the electron density participate in the determination
of the cavity, since the nearest approach of closed shell systems only involves overlap
of these low electron density regions. It turns out that we can exploit this exponential
structure to much more reliably expand the electron density rather than the cavities
obtained by thresholding them.

Isodensity PCM's describe the cavities by functions $s(\vec{r})$ that smoothly switch
from 0 in the cavity to 1 in the bulk fluid (the corresponding spatially-varying
dielectric constant is $\epsilon(\vec{r}) = 1 + (\epsilon_b-1) s(\vec{r})$).
Following Refs.~\citenum{JDFT} and \citenum{NonlinearPCM}, we employ the error-function form
\begin{equation}
s_R[n](\vec{r}) = \frac{1}{2}\textrm{erfc} \frac{\ln(\eta_R[n]/n_c)}{\sigma\sqrt{2}}, \label{eqn:ShapeExpanded}
\end{equation}
with $\sigma = 0.6$ selected such that the sharpness of the transition from 0 to 1
for typical electron densities is resolvable on plane-wave grids at typical kinetic energy cutoffs.
The parameter $n_c$ will be determined by fits to solvation energies in section~\ref{sec:SolvationEnergies}.
The key difference here is that we use $\eta_R[n]$, the electron density expanded by $R$,
instead of the original electron density $n$ to obtain an expanded cavity.
In particular, according to the program of figure~\ref{fig:CavityRelations},
the electrostatic cavity is $s\sub{el} \equiv s_{R\sub{vdW} - R\sub{el}}$
and the solvent-center cavity is $s\sub{SAS} \equiv s_{R\sub{vdW}}$.

Finally, we specify the functional $\eta_R[n]$ that expands
the exponential tails of the electron density.
Convolving the electron density by a weight function with range $R$
almost achieves the required task, since the result at any location
is dominated by the highest electron density within the range,
which would be from $R$ `inwards' from that location.
In particular, with a spherical kernel $w(r) = \theta(R-r)/2\pi R^3$
(with a convenient dimensionless normalization), a planar electron
density $n = \exp(-z/a)$ yields a convolved density 
\begin{equation}
\bar{n} = \frac{a^2(R-a)}{R^3}e^{-(z-R)/a} + \mathcal{O}\left(e^{-(z+R)/a}\right),
\end{equation}
which exhibits the desired shifting of the exponential tail
but includes an undesirable prefactor that depends on the electron density length scale, $a$.
A gradient of the convolution picks up a factor of $1/a$, and can be combined with
the above to eliminate this dependence. In fact, $|R\nabla\bar{n}|^2/\bar{n}
= \exp(-(z-R)/a)) + \mathcal{O}(a/R)$ and $R \gg a$ for typical electron densities.
This form, however, rapidly approaches zero in the core region of pseudized
electron densities. A sum of the convolved density, $\bar{n} = n \ast \theta(R-r)/2\pi R^3$,
and the combination with its gradient,
\begin{equation}
\eta_R[n] = \bar{n} + \frac{|R\nabla\bar{n}|^2}{\bar{n}} \label{eqn:nExpansion}
\end{equation}
retains the leading order electron density length-scale independence and remains
non-zero in the pseudopotential cores. Figure~\ref{fig:nExpansion} demonstrates
the accuracy of this functional in expanding realistic electron densities
and the resulting cavities. The errors in cavity separation for planar electron densities,
the regime of the above construction, are only $\sim 0.01$~\Angstrom,
whereas they approach $\sim 0.03$~\Angstrom~ for worst-case spherical densities
with curvature radii comparable to the expansion radius $R$.
Typical separations between solute and solvent atoms, which do not form covalent bonds
with each other, are $\sim 3$~\Angstrom, so that the worst-case error above results
in a 1~\% error in the capacitance of the dielectric cavity, and hence a 1~\% error
in the electrostatic contribution to the solvation energy. This component of the energy
typically varies from $\sim 10$~m$E_h$ for organic molecules to $\sim 100$~m$E_h$
for ions, so that the worst-case error would be $\sim 1$~m$E_h$ (room temperature) for ions,
which is within the target accuracy of $\sim 1$~kcal/mol (1.6~m$E_h$) for simplified solvation models.
 
\makeatletter{}\section{Weighted-density cavity formation model} \label{sec:CavitationModel}

The analyses of Sections~\ref{sec:ElectrostaticRadius} and \ref{sec:CavityExpansion}
establish a solvent-center cavity and a dielectric cavity within an iso-density
approach with a single critical electron density parameter $n_c$. 
The remaining ingredients necessary to form a complete polarizable continuum description
of the solvent are models for the sub-dominant contributions beyond the mean-field
electrostatic interactions, such as the cavity formation and dispersion energies,
given the solvent-center cavity described in term of the shape function, $s\sub{SAS}(\vec{r})$.

The simplest approximation to the cavity formation free energy is an effective
surface tension model, with an empirical tension parameter fit to solvation free energies.
This empirically accounts for the reduced free energy per unit surface relative to the bulk
surface tension for microscopic molecular cavities,\cite{HardSphereSPCE} but therefore underestimates
the cavity contribution for a planar interface, which should in fact be the bulk surface tension.
A model accounting for the cavity geometry is necessary to describe both limits accurately.

The cavities in traditional polarizable continuum models are typically composed of spheres.
Scaled-particle theory (SPT),\cite{SPT-Review} based on the statistical mechanics of hard sphere fluids,
provides an accurate estimate of the free energy of inserting a hard sphere of arbitrary size
into a hard sphere fluid. PCM's employ various combination rules \cite{PCM-Review} to estimate 
the free energy for forming a cavity composed as a union of spheres, such as applying SPT
to a sphere of surface area or volume equal to that of the cavity, or linearly combining
the cavity formation energy of spheres weighted by exposed surface area.
(See Ref.~\citenum{Cavitation-Compare} for a detailed comparison of these methods.)

These combination rules do not result from physical principles and have primarily evolved
from empirical evidence. Furthermore, isodensity PCM's produce arbitrary-shaped cavities
that do not decompose into spheres for which SPT may be applied.
In principle, classical density functional theory with free energy functional approximations\cite{BondedTrimer,RigidCDFT,PolarizableCDFT} can provide an estimate of this term.
This involves minimizing a free energy functional in an external potential
that excludes the fluid from the interior of the cavity, which incurs a significant computational cost
compared to the solution of the modified dielectric equation for the electrostatic term. 
Here, we motivate a low computational cost, closed-form physical model for the cavity-formation energy
for arbitrary cavities that compares favorably with classical-density functional results.

We start from the intuitive picture of surface tension resulting from the energy cost
of missing neighbors for the molecules at the surface of the fluid.
A convolution of the cavity shape function, $\bar{s} = w \ast s$, with a normalized
short-ranged weight function $w(r)$, measures the neighborhood of a molecule, ranging from 0
for an isolated molecule, through $1/2$ for a surface molecule, to 1 for a molecule in the bulk.
In particular, we select the spherical shell weight function
$w(r) = \delta(r-\sigma\sub{vdW})/4\pi \sigma\sub{vdW}^2$ with the solvent vdW diameter $\sigma\sub{vdW}=2R\sub{vdW}$,
so as to estimate the fraction of nearest neighbor molecules present at each location.
We then make a weighted-density ansatz for the cavity formation free energy,
$G\sub{cav} = pV + \int d\vec{r} f(\bar{s})$ with an as yet undetermined local function $f$,
after separating out the ideal gas contribution $pV$ for a cavity of volume $V = \int (1-s)$
in a fluid at pressure $p$ and temperature $T$.

Next, we constrain the undetermined function to known physical limits.
The free energy to form a cavity of volume $V$ that is much smaller than molecular dimensions
in a fluid of bulk density $N\sub{bulk}$ at temperature $T$, is dominated by ideal gas contributions
and reduces to $(p + N\sub{bulk}T) V$ to lowest order in $V$.
On the other hand, the weighted density ansatz above yields $G\sub{cav} = f(1) + (f'(1)+p)V + \mathcal{O}(V^2)$
in the limit of small cavities, which implies $f(1) = 0$ and $f'(1) = N\sub{bulk}T$.

The opposite regime of droplets corresponds to fluid at bulk density in the interior of some volume $V$,
with zero density outside. When we take the limit $V \rightarrow 0$, this configuration contains
$N\sub{bulk}V \ll 1$ molecules on average (no longer a droplet in the conventional sense),
and its free energy corresponds to that of extracting and isolating $N\sub{bulk}V$ molecules from the bulk fluid.
The free energy required to isolate one molecule from the bulk fluid is related to its vapor pressure, $p\sub{vap}$,
as $T (\ln \frac{N\sub{bulk}T}{p\sub{vap}}-1)$. In this limit, the ansatz predicts
$G\sub{cav} = f(0) + f'(0)V + \mathcal{O}(V^2)$, yielding the constraints $f(0) = 0$
and $f'(0) = N\sub{bulk}T (\ln \frac{N\sub{bulk}T}{p\sub{vap}}-1)$.

In addition to these four constraints from the droplet and cavity limits,
we require the model to reproduce the bulk surface tension $\sigma\sub{bulk}$ for planar interfaces.
The simplest function $f(\bar{s})$ that can satisfy these five constraints is a fourth order polynomial,
and solving for the constraints for that functional form yields the model free energy
\begin{multline}
G\sub{cav}[s] = p\int(1-\bar{s}) + N\sub{bulk}T \int \bar{s}(1-\bar{s}) \bigg[ \bar{s} + (1-\bar{s})\gamma \\
	\left. + 15\bar{s}(1-\bar{s}) \left( \frac{\sigma\sub{bulk}}{N\sub{bulk}TR\sub{vdW}} - \frac{1+\gamma}{6} \right) \right],
\label{eqn:CavitationModel}
\end{multline}
where $\gamma \equiv \ln \frac{N\sub{bulk}T}{p\sub{vap}}-1$.

\begin{figure}
\includegraphics[width=\columnwidth]{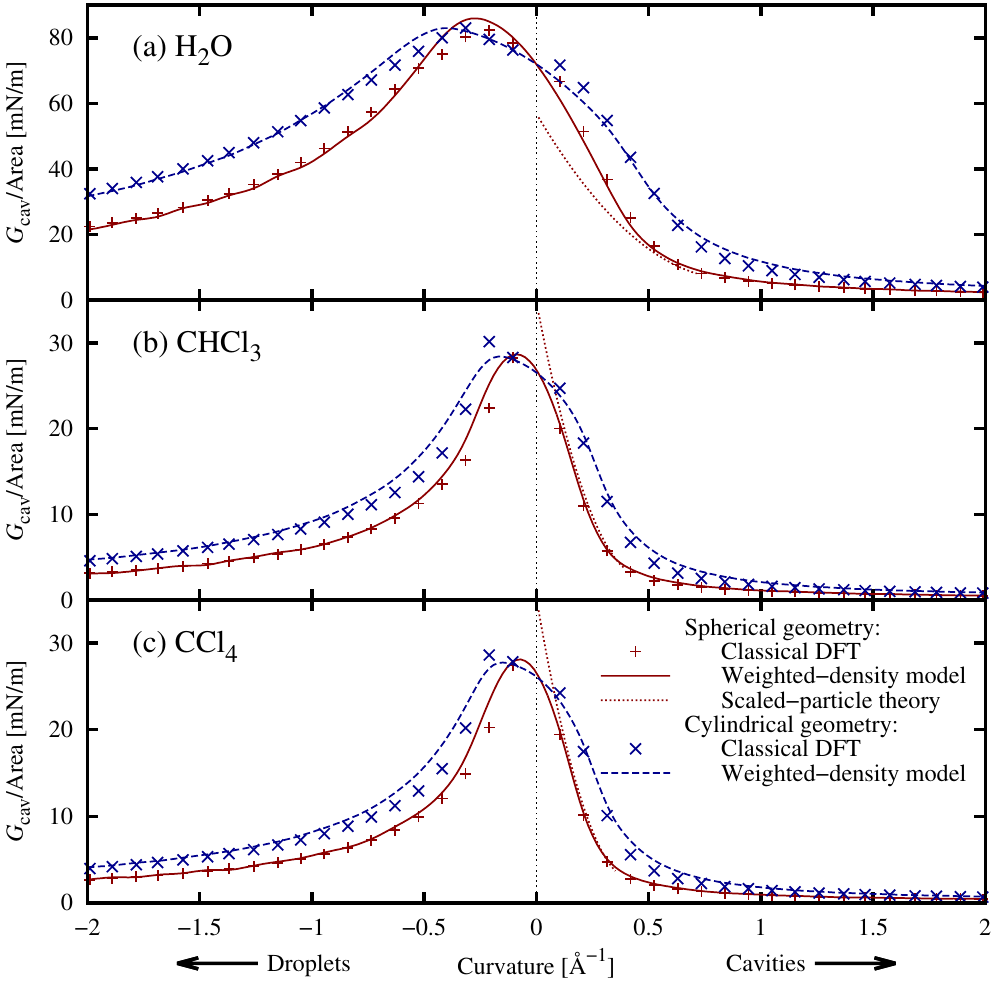}
\caption[Dependence on the fluid-vacuum interface curvature of the cavity formation surface energy]
{Dependence on the curvature of the fluid-vacuum interface,
of the cavity formation energy per surface area
for (a) water, (b) chloroform and (c) carbon tetrachloride
as predicted by the weighted density model (\ref{eqn:CavitationModel})
and scaled-particle theory \cite{SPT-Review} (valid for spherical cavities only)
compared to classical density functional results for spherical
and cylindrical cavities (fluid outside surface) as well as droplets
(fluid inside surface).
\label{fig:sigmaVsCurvature}}
\end{figure}

\begin{figure}
\includegraphics[width=\columnwidth]{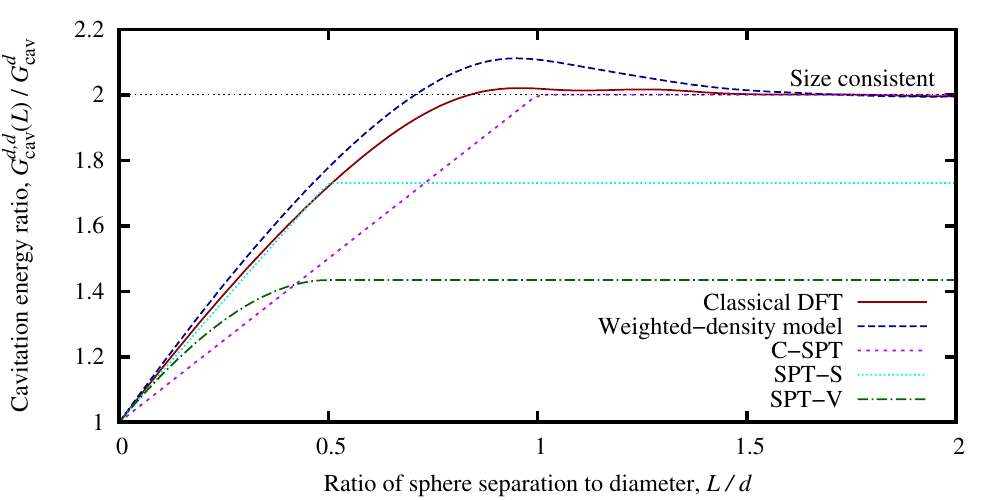}
\caption[Size consistency of various cavity formation energy models]
{Ratio of cavity formation energies for a union of two spherical solvent-center cavities
of diameter $d$ ($G\sub{cav}^{d,d}(L)$) to that of a single isolated cavity
($G\sub{cav}^{d}$) as a function of sphere center separation $L$,
as predicted by the weighted density model and several traditional
PCM combination rules compared to classical density functional results.
Zero separation corresponds to a single spherical cavity, while infinite
separation corresponds to two isolated spherical cavities.
A size-consistent model must yield 2 in the limit of infinite separation.
The figure presents results for water, with solute spheres of radius
equal to the solvent $R\sub{vdW}$, which results in solvent-center
cavities of diameter $d=4R\sub{vdW}=5.54$~\Angstrom, and
for which the isolated cavity formation energy is $G\sub{cav}^{d} = 4.3$~kcal/mol (6.9~m$E_h$).
\label{fig:sigmaSizeConsistency}}
\end{figure}

This specifies a model for the free energy associated with forming a cavity of arbitrary shape
described by a shape function $s(\vec{r})$, constrained entirely by bulk
measurable properties of the solvent, with no adjustable parameters.
Figure~\ref{fig:sigmaVsCurvature} compares the predictions of this model
against classical density functional theory calculations using the free energy functional
approximations of Ref.~\citenum{PolarizableCDFT} for all constant curvature surfaces:
spherical and cylindrical cavities as well as droplets. The model reproduces
the high positive and negative as well as zero curvature results by construction
(small cavity, small droplets and planar interface limits respectively).
The weighted density ansatz with the spherical shell weight function
of radius $R\sub{vdW}$ perfectly `interpolates' between these limits
for all solvents considered, ranging from the highly polar (water)
to the non-polar (carbon tetrachloride). The results are also in agreement
with scaled-particle theory in the latter's domain of validity: small spherical cavities,
but the present model is valid for arbitrary geometries and does not require
combination rules for application to PCM free energies for solvated molecules.

Figure~\ref{fig:sigmaSizeConsistency} further explores the accuracy of this model for
non-spherical geometries by considering the cavity formation energy for dumbbell-shaped
objects composed as the union of two spherical cavities, as a function of the
separation between the sphere centers. The weighted-density model best reproduces
the classical density-functional results, including the non-monotonicity around
separations for which the two cavities just touch; this corresponds to a highly
non-analytical geometry involving cusps and infinite surface curvatures.
Further, the minor deviations from the density-functional results are only for
this problematic non-analytical geometry. On the other hand, amongst the traditional PCM
combination rules, only the Claviere-Peirotti method (C-SPT) \cite{Cavitation-CSPT} that
combines the sphere results weighted by the exposed surface areas exhibits size-consistency,
that is it evaluates to twice the cavity formation energy for two infinitely separated
spherical cavities, compared to that for a single cavity.
The other methods that apply scaled-particle theory to a sphere with either
surface area or volume equal to the non-spherical cavity (SPT-S and SPT-V respectively \cite{Cavitation-Compare})
sacrifice the size consistency, but are more accurate for small separations
where the resulting cavity approaches a sphere.
Clearly, as expected, the weighted density model consistently exhibits the best results
for highly non-spherical geometries.
 
\makeatletter{}\section{Dispersion model} \label{sec:DispersionModel}

The final energetic contributions relevant to solvation free energies
are the dispersion interactions between the solute and the solvent,
and to a lesser extent, the Pauli repulsion between the electrons
of the solute and the solvent at their interface.
Quantum chemistry solvation methods sometimes couple solvent polarizabilities
to virtual excitations in the solute system to obtain a physical model
for the dispersion interactions;\cite{RepulsionDispersionModel}
such methods are much more expensive than standard electronic density-functional
calculations, require unoccupied levels and can be prohibitively expensive
in plane-wave basis calculations. On the other hand, empirical pair-potential
estimates for these additional terms \cite{PCM-Review} are moderately accurate
and efficient for use in density-functional calculations.

Here, we neglect the repulsion energies and adopt a simplified empirical formulation
for the dispersion energies (which absorbs the error introduced by neglecting repulsion)
based on the pair-potential dispersion corrections employed in electronic
density-functional theory.\cite{Dispersion-Grimme,Dispersion-TS}
For simplicity, we adapt Grimme's form,\cite{Dispersion-Grimme}
which expresses the dispersion corrections for a system
with a collection of atoms at positions $\vec{R}_i$ as
\begin{equation}
E\sub{disp} = -s_6 \sum_{i<j} \frac{\sqrt{C_{6i}C_{6j}}}{r_{ij}^6}
	f\sub{dmp}\left(\frac{r_{ij}}{R_{0i}+R_{0j}}\right),
\end{equation}
where $r_{ij} \equiv |\vec{R}_i-\vec{R}_j|$, $C_{6i}$ are effective interaction
strengths for each atom type derived from \emph{ab initio} atomic polarizabilities,
and $R_{0i}$ are atomic vdW radii (tabulated for all main group elements in \cite{Dispersion-Grimme}).
The damping function $f\sub{dmp}(x) = 1/(1+e^{-d(x-1)})$ with $d=23$ serves to
attenuate the short-ranged contributions to the correction, since they are
partially captured by the approximate exchange-correlation functional,
and the empirical scale parameter $s_6$ compensates for differences
between various exchange-correlation functionals.

We make two modifications in adapting this pair-potential model for solvent-solute
interactions in PCM. First, the damped $r^{-6}$ potential is still singular at
zero separation and not integrable ($\int dx 4\pi x^2 f\sub{dmp}(x)/x^6 = \infty$).
This makes no difference since atoms never get close enough for this
unphysical behavior to contribute, but the lack of integrability precludes evaluating the
interaction with continuous distributions of atoms using a convolution. Therefore, we eliminate the
$x=0$ singularity and instead employ the value and derivative matched piecewise function
\begin{equation}
f\sub{dmp}(x) = \begin{cases}
	 1/(1+e^{-d(x-1)}), & x > 0.03\\
	0.00114 x^6, &  x \leq 0.03
\end{cases}
\end{equation}
which is identical to the original function at all relevant distances.

Second, the simplified PCM description of the solvent does not specify
spatial distributions for each atom of the solvent molecule,
but only the distribution of the solvent molecule centers, $N\sub{bulk}s(\vec{r})$.
Consequently, we additionally assume a uniform orientation distribution
of the molecules comprising the cavity resulting in a spatial distribution
$N_j(\vec{r}) = N\sub{bulk}s(\vec{r}) \ast \delta(r-R_j\super{solv})/4\pi (R_j\super{solv})^2$
for atom $j$ of the solvent molecule that is at a distance $R_j\super{solv}$ from the
center of the solvent molecule. This results in a model solvent-solute dispersion interaction
\begin{multline}
E\sub{disp}[s] = -s_6 \sum_{i,j} \int d\vec{r} N_j(\vec{r})
	\frac{\sqrt{C_{6i}C_{6j}}}{|\vec{R}_i-\vec{r}|^6} \\ \times
	f\sub{dmp}\left(\frac{|\vec{R}_i-\vec{r}|}{R_{0i}+R_{0j}}\right), \label{eqn:DispersionModel}
\end{multline}
where the index $i$ runs over the atoms of the explicit electronic system
and $j$ over the atoms of one solvent molecule.
The empirical scale factor $s_6$ absorbs the errors arising from the neglect of
repulsion as well as the uniform orientation distribution assumption,
in addition to those inherent to the pair-potential approximation,
such as the neglect of three-body (Axilrod-Teller) terms and beyond.

The disadvantage of this simplified description is the introduction of one solvent-dependent
empirical parameter, which we mitigate in the following fits to solvation energies.
In particular, we show that this single parameter can be calibrated to the solvation energy
of a single non-polar molecule which is dominated by the dispersion interaction.
Along with the solvent-independent $n_c$, this allows the application of the method
to an arbitrary solvent without requiring extensive fits to solvation energy data sets.
 
\makeatletter{}\section{Solvation energies} \label{sec:SolvationEnergies}

The previous sections establish the relations between the cavity for electric response and
that of the solvent centers, and formulate weighted-density models for the cavity formation
and dispersion energies that capture the true shape and size dependence of these contributions.
Combining these with the nonlinear electric and ionic response of Ref.~\citenum{NonlinearPCM},
we arrive at the modified nonlinear solvation contribution,
\begin{multline}
A\sub{diel} = A_\epsilon[s\sub{el},\vec{\varepsilon}] + A_\kappa[s\sub{el},\eta_\pm] \\
	+ \int d\vec{r} \int d\vec{r}' \frac{\rho\sub{lq}(\vec{r}')}{|\vec{r}-\vec{r}'|}
		\left(\rho\sub{el}(\vec{r}) + \frac{\rho\sub{lq}(\vec{r})}{2} \right) \\
	+ G\sub{cav}[s\sub{SAS}] + E\sub{disp}[s\sub{SAS}].
\label{eqn:modifiedPCM}
\end{multline}
The first term represents the internal energy functional of a continuum dielectric with a local
but non-linear response, expressed in terms of the independent variable $\varepsilon(\vec{r})$
which corresponds to the effective local electric field. The second optional term represents
the internal energy contribution due to ions in the solution (if any), expressed in terms
of the local chemical potentials for the anions and cations $\eta_\pm(\vec{r})$.
The third term accounts for the mean field interactions of the bound charge in the liquid
$\rho\sub{lq}(\vec{r})$ (which includes dielectric and optionally ionic contributions) with itself
and with the total charge density (electronic and ionic) $\rho\sub{el}(\vec{r})$ of the solute system.
The last two terms account for the cavity formation and dispersion energies as detailed
in sections~\ref{sec:CavitationModel} and \ref{sec:DispersionModel}.
The free energy $A\sub{diel}$ is added to that of a standard Kohn-Sham electronic density functional
to obtain the free energy including liquid effects, and the net free energy is minimized self-consistently.
See Ref.~\citenum{NonlinearPCM} for a detailed description of the internal energy terms,
algorithms for minimizing the fluid free energy and the implementation of the overall framework in JDFTx.\cite{JDFTx}

The difference here in contrast to Ref.~\citenum{NonlinearPCM},
is that the first two terms of (\ref{eqn:modifiedPCM})
employ the electrostatic cavity $s\sub{el}(\vec{r})$,
while the final two terms, for cavity formation and dispersion,
employ physical models evaluated on the solvent-center cavity,
$s\sub{SAS}(\vec{r})$ (instead of a combined empirical surface tension model).
The solute electron density determines both cavities with
a single critical density, $n_c$, according to (\ref{eqn:ShapeExpanded}).
Bulk solvent properties and \emph{ab initio} calculations on a single solvent molecule
determine all terms in this model, except $n_c$ and the dispersion scale factor, $s_6$.

\begin{figure}
{\large (a) H\sub{2}O}\\
\includegraphics[width=\columnwidth]{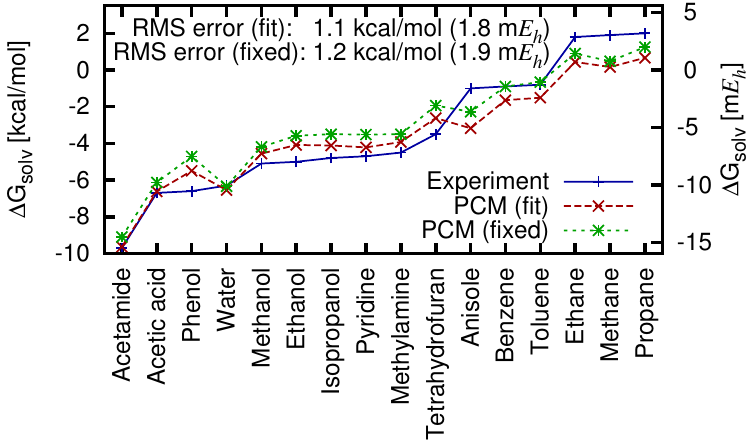} \\
{\large (b) CHCl\sub{3}} \\
\includegraphics[width=\columnwidth]{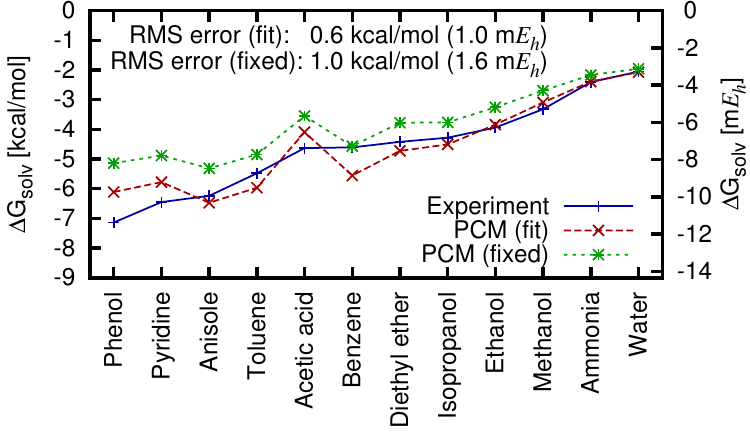} \\
{\large (c) CCl\sub{4}} \\
\includegraphics[width=\columnwidth]{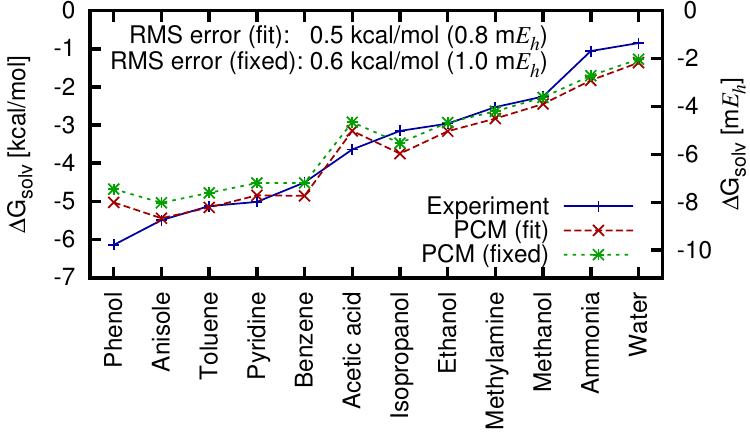}
\caption[Solvation energies predicted by nonlinear PCM with weighted-density cavity formation and dispersion terms]
{Solvation energies predicted by nonlinear PCM with weighted-density cavity formation and dispersion terms
for molecules in (a) water, (b) chloroform and (c) carbon tetrachloride, compared to experiment.
Phenol and acetic acid seem to be outliers in (b) and (c), probably due to a weak chemical
bond between the solute and solvent in the experiment that is missed by the solvation model.
\label{fig:CavWDA_SolvationEnergies}}
\end{figure}

The critical density $n_c$ corresponds to the base cavity in Figure~\ref{fig:CavityRelations},
which we expect to be a property of the solute alone and hence solvent-independent.
(The solvent-dependence of the electrostatic cavity size is due to the nonlocality of the
true response of the solvent; this effect enters the calculation of the electrostatic radius in
Section~\ref{sec:ElectrostaticRadius} and we no longer expect it to affect the base cavity size and $n_c$.)
The dispersion scale factor, $s_6$, absorbs errors due to the neglect of repulsion terms
and the assumption of isotropic solvent distribution in (\ref{eqn:DispersionModel}),
and may depend on the solvent. We therefore fit a single $n_c$ and an $s_6$ per solvent to the
solvation energies of several small molecules in water, chloroform and carbon tetrachloride.
Table~\ref{tab:fitParamsSGA} shows the resulting fit parameters and residuals, and
Figure~\ref{fig:CavWDA_SolvationEnergies} shows the solvation energies of each molecule
compared to experimental data.\cite{solvation-exp-compiled-1,solvation-exp-compiled-2}

\begin{table}
\caption{Fit parameters and residuals for nonlinear PCM with weighted-density cavity formation and dispersion terms
\label{tab:fitParamsSGA}}
\begin{center}\begin{tabular}{cc|cc|cc}
\hline\hline
\multirow{2}{*}{Solvent} & \multirow{2}{*}{$n_c\ [a_0^{-3}]$} & \multicolumn{2}{c|}{$s_6$} & \multicolumn{2}{c}{\parbox{1in}{RMS error\\{[}kcal/mol (m$E_h$)]}} \\
& & Fit & Fixed & Fit & Fixed \\
\hline
H\sub{2}O   & \multirow{3}{*}{\sci{1.0}{-2}} & 0.54 & 0.50 & 1.1 (1.8) & 1.2 (1.9)\\
CHCl\sub{3} &                                & 1.31 & 1.08 & 0.6 (1.0) & 1.0 (1.6) \\
CCl\sub{4}  &                                & 1.24 & 1.20 & 0.5 (0.8) & 0.6 (1.0) \\
\hline\hline
\end{tabular}\end{center}
\end{table}

\begin{table}
\caption{Fit parameters and residuals for nonlinear PCM with empirical cavity surface tension
\label{tab:fitParamsGLSSA}}
\begin{center}\begin{tabular}{cccc}
\hline\hline
Solvent & $n_c\ [a_0^{-3}]$ & $\tau\ [E_h/a_0^2]$ & \parbox{1in}{RMS error\\{[}kcal/mol (m$E_h$)]} \\
\hline
H\sub{2}O    & \sci{1.0}{-3} & \sci{9.5}{-6}  & 1.0 (1.6)\\
CHCl\sub{3}  & \sci{2.4}{-5} & \sci{-9.2}{-6} & 0.8 (1.3)\\
CCl\sub{4}   & \sci{1.2}{-4} & \sci{-9.0}{-6} & 1.1 (1.8)\\
\hline\hline
\end{tabular}\end{center}
\end{table}

A single $n_c$ indeed fits the solvation energies of vastly different solvents, ranging
from the small polar water, to the much larger non-polar carbon tetrachloride.
The dispersion factor $s_6$ varies between the solvents, but remains within 35\% of the
range 0.75-1.2 covered by different electronic functionals in \cite{Dispersion-Grimme}.

In contrast, the original nonlinear PCM of Ref.~\citenum{NonlinearPCM}, which employs
an empirical surface tension $\tau$ on the surface of the electric response cavity
to account for both cavity formation and dispersion, requires wildly different $n_c$'s
for the three solvents, covering three orders of magnitude, as Table~\ref{tab:fitParamsGLSSA} shows.
Further, the effective surface tensions for the less polar solvents, chloroform and carbon tetrachloride, are
negative since the attractive dispersion effects dominate over the repulsive cavity formation energies.
This negative tension contributes a strong attractive well to the electron potential, which occasionally
renders the electronic density functional unstable with respect to leaking electrons into the cavity.
Additionally, the $n_c$ for these solvents is much smaller on account of the larger molecular size,
making the calculations more sensitive to the Nyquist frequency noise in the electron density.
It is therefore advisable to avoid the effective surface tension
approach\cite{PCM-SCCS,PCM-SCCS-charged,NonlinearPCM,PCM-correlateTau}
for non-polar dispersion-dominated solvents.

More importantly, the current model with physical cavity formation and dispersion terms
obtains the same accuracy as the previous models, and does so with fewer parameters.
Indeed, Table~\ref{tab:fitParamsSGA} shows that the fit results in $\sim 1$ kcal/mol (1.6~m$E_h$) accuracy
for all three solvents, which is sufficient for the study of chemical reactions in solution.

Finally, the use of a common $n_c$ for all solvents makes it easier to extend the method to other solvents,
requiring the determination of a single fit parameter $s_6$ for each new solvent. In the absence of extensive data
for a new solvent, it should be possible to use the solvation energy of just \emph{one} solute
to calibrate this single parameter. The ideal molecule for this purpose should be large, polarizable and
non-polar, so that dispersion interactions dominate its solvation energy and constrain $s_6$ reliably.
Figure~\ref{fig:CavWDA_SolvationEnergies} shows that fixing $s_6$ to reproduce the solvation energy of benzene
in each solvent only marginally worsens the errors relative to the full fits. The resulting `fixed' $s_6$
parameters are similar to the fit ones, and the residual remains in the $\sim 1$ kcal/mol (1.6 m$E_h$) regime,
as Table~\ref{tab:fitParamsSGA} shows.

\section{Conclusions}

This work analyzes the distinct atom-based and density-based parametrizations
of continuum solvation models, combines the best features from these two
distinct approaches, and introduces physical models for the cavity formation
and dispersion energies, to construct a minimally-empirical solvation model.
The key idea is to follow the traditional PCM approach of using a dielectric
cavity distinct from (and smaller than) the solvent-center surface (SAS),
but to instead determine both cavities from the electron density using
a single universal parameter $n_c$, as explained in section~\ref{sec:CavityExpansion}.
Additionally, we compute the separation $R\sub{el}$ between these two cavities
from the nonlocal response of a one dimensional solvent vacuum interface
according to the \emph{ansatz} of section~\ref{sec:ElectrostaticRadius},
rather than fitting it to solvation energies.

This work also presents a closed-form weighted density approximation
(section~\ref{sec:CavitationModel}) to the free energy of forming a cavity of
arbitrary shape and size in the liquid, which agrees well with classical
density functional theory calculations (which are computationally much more expensive).
Along with a pair-potential model for the dispersion interactions
(section~\ref{sec:DispersionModel}), this allows us to capture
the correct cavity size and shape dependence of these additional terms,
instead of making an unphysical surface area or volume dependent approximation.

The resulting model has one solvent-independent parameter $n_c$ and one
parameter per solvent, the dispersion scale factor $s_6$. We fit one $n_c$
and three $s_6$'s (one per solvent) to solvation energies of several
organic molecules in water, chloroform and carbon tetrachloride,
and demonstrate that we obtain an RMS error around 1 kcal/mol,
which is comparable to that of previous density-based solvation models
with two or more parameters per solvent.\cite{NonlinearPCM,PCM-SCCS}
We also show that the $s_6$ can be estimated from the solvation energy
of a single non-polar molecule, without sacrificing much accuracy,
thereby further easing the parametrization of this model for new solvents.

Finally, this work opens up two avenues for future work.
First, further theoretical developments towards the nonlocal dielectric response
could improve upon the electrostatic radii and perhaps the universal parameter $n_c$
could be estimated independently without reference to solvation energies.
This could lead to a model with just a single parameter per solvent.
Second, in a complimentary direction, we could treat $R\sub{el}$ as an additional
fit parameter, and optimize it to improve the accuracy of the model; the estimate
of section~\ref{sec:ElectrostaticRadius} would then serve as an educated starting guess.
This might be necessary for solvents with larger molecules, where the rigid rotor
approximation would no longer be valid.

This work was supported as a part of the Energy Materials Center at Cornell (EMC$^2$),
an Energy Frontier Research Center funded by the U.S. Department of Energy,
Office of Science, Office of Basic Energy Sciences under Award Number DE-SC0001086.
Additional computational time at the Texas Advanced Computing Center (TACC) at the University of Texas
at Austin, was provided via the Extreme Science and Engineering Discovery Environment (XSEDE),
which is supported by National Science Foundation grant number OCI-1053575.

\makeatletter{} 


\begin{thebibliography}{27}\makeatletter
\providecommand \@ifxundefined [1]{ \@ifx{#1\undefined}
}\providecommand \@ifnum [1]{ \ifnum #1\expandafter \@firstoftwo
 \else \expandafter \@secondoftwo
 \fi
}\providecommand \@ifx [1]{ \ifx #1\expandafter \@firstoftwo
 \else \expandafter \@secondoftwo
 \fi
}\providecommand \natexlab [1]{#1}\providecommand \enquote  [1]{``#1''}\providecommand \bibnamefont  [1]{#1}\providecommand \bibfnamefont [1]{#1}\providecommand \citenamefont [1]{#1}\providecommand \href@noop [0]{\@secondoftwo}\providecommand \href [0]{\begingroup \@sanitize@url \@href}\providecommand \@href[1]{\@@startlink{#1}\@@href}\providecommand \@@href[1]{\endgroup#1\@@endlink}\providecommand \@sanitize@url [0]{\catcode `\\12\catcode `\$12\catcode
  `\&12\catcode `\#12\catcode `\^12\catcode `\_12\catcode `\%12\relax}\providecommand \@@startlink[1]{}\providecommand \@@endlink[0]{}\providecommand \url  [0]{\begingroup\@sanitize@url \@url }\providecommand \@url [1]{\endgroup\@href {#1}{\urlprefix }}\providecommand \urlprefix  [0]{URL }\providecommand \Eprint [0]{\href }\providecommand \doibase [0]{http://dx.doi.org/}\providecommand \selectlanguage [0]{\@gobble}\providecommand \bibinfo  [0]{\@secondoftwo}\providecommand \bibfield  [0]{\@secondoftwo}\providecommand \translation [1]{[#1]}\providecommand \BibitemOpen [0]{}\providecommand \bibitemStop [0]{}\providecommand \bibitemNoStop [0]{.\EOS\space}\providecommand \EOS [0]{\spacefactor3000\relax}\providecommand \BibitemShut  [1]{\csname bibitem#1\endcsname}\let\auto@bib@innerbib\@empty
\bibitem [{\citenamefont {Fortunelli}\ and\ \citenamefont
  {Tomasi}(1994)}]{PCM94}  \BibitemOpen
  \bibfield  {author} {\bibinfo {author} {\bibfnamefont {A.}~\bibnamefont
  {Fortunelli}}\ and\ \bibinfo {author} {\bibfnamefont {J.}~\bibnamefont
  {Tomasi}},\ }\href@noop {} {\bibfield  {journal} {\bibinfo  {journal} {Chem.
  Phys. Lett.}\ }\textbf {\bibinfo {volume} {231}},\ \bibinfo {pages} {34}
  (\bibinfo {year} {1994})}\BibitemShut {NoStop}\bibitem [{\citenamefont {Barone}, \citenamefont {Cossi},\ and\ \citenamefont
  {Tomasi}(1997)}]{PCM97}  \BibitemOpen
  \bibfield  {author} {\bibinfo {author} {\bibfnamefont {V.}~\bibnamefont
  {Barone}}, \bibinfo {author} {\bibfnamefont {M.}~\bibnamefont {Cossi}}, \
  and\ \bibinfo {author} {\bibfnamefont {J.}~\bibnamefont {Tomasi}},\
  }\href@noop {} {\bibfield  {journal} {\bibinfo  {journal} {J. Chem. Phys.}\
  }\textbf {\bibinfo {volume} {107}},\ \bibinfo {pages} {3210} (\bibinfo {year}
  {1997})}\BibitemShut {NoStop}\bibitem [{\citenamefont {Tomasi}, \citenamefont {Mennucci},\ and\
  \citenamefont {Cammi}(2005)}]{PCM-Review}  \BibitemOpen
  \bibfield  {author} {\bibinfo {author} {\bibfnamefont {J.}~\bibnamefont
  {Tomasi}}, \bibinfo {author} {\bibfnamefont {B.}~\bibnamefont {Mennucci}}, \
  and\ \bibinfo {author} {\bibfnamefont {R.}~\bibnamefont {Cammi}},\
  }\href@noop {} {\bibfield  {journal} {\bibinfo  {journal} {Chem. Rev.}\
  }\textbf {\bibinfo {volume} {105}},\ \bibinfo {pages} {2999} (\bibinfo {year}
  {2005})}\BibitemShut {NoStop}\bibitem [{\citenamefont {Cramer}\ and\ \citenamefont
  {Truhlar}(1991)}]{PCM-SM1}  \BibitemOpen
  \bibfield  {author} {\bibinfo {author} {\bibfnamefont {C.~J.}\ \bibnamefont
  {Cramer}}\ and\ \bibinfo {author} {\bibfnamefont {D.~G.}\ \bibnamefont
  {Truhlar}},\ }\href@noop {} {\bibfield  {journal} {\bibinfo  {journal} {J.Am.
  Chem. Soc.}\ }\textbf {\bibinfo {volume} {113}},\ \bibinfo {pages} {8305}
  (\bibinfo {year} {1991})}\BibitemShut {NoStop}\bibitem [{\citenamefont {Marenich}\ \emph {et~al.}(2007)\citenamefont
  {Marenich}, \citenamefont {Olson}, \citenamefont {Kelly}, \citenamefont
  {Cramer},\ and\ \citenamefont {Truhlar}}]{PCM-SM8}  \BibitemOpen
  \bibfield  {author} {\bibinfo {author} {\bibfnamefont {A.~V.}\ \bibnamefont
  {Marenich}}, \bibinfo {author} {\bibfnamefont {R.~M.}\ \bibnamefont {Olson}},
  \bibinfo {author} {\bibfnamefont {C.~P.}\ \bibnamefont {Kelly}}, \bibinfo
  {author} {\bibfnamefont {C.~J.}\ \bibnamefont {Cramer}}, \ and\ \bibinfo
  {author} {\bibfnamefont {D.~G.}\ \bibnamefont {Truhlar}},\ }\href@noop {}
  {\bibfield  {journal} {\bibinfo  {journal} {J. Chem. Theory Comput.}\ ,\
  \bibinfo {pages} {2011}} (\bibinfo {year} {2007})}\BibitemShut {NoStop}\bibitem [{\citenamefont {Marenich}, \citenamefont {Cramer},\ and\
  \citenamefont {Truhlar}(2009)}]{PCM-SMD}  \BibitemOpen
  \bibfield  {author} {\bibinfo {author} {\bibfnamefont {A.~V.}\ \bibnamefont
  {Marenich}}, \bibinfo {author} {\bibfnamefont {C.~J.}\ \bibnamefont
  {Cramer}}, \ and\ \bibinfo {author} {\bibfnamefont {D.~G.}\ \bibnamefont
  {Truhlar}},\ }\href@noop {} {\bibfield  {journal} {\bibinfo  {journal} {J.
  Phys. Chem. B}\ }\textbf {\bibinfo {volume} {113}},\ \bibinfo {pages} {6378}
  (\bibinfo {year} {2009})}\BibitemShut {NoStop}\bibitem [{\citenamefont {Andreussi}, \citenamefont {Dabo},\ and\ \citenamefont
  {Marzari}(2012)}]{PCM-SCCS}  \BibitemOpen
  \bibfield  {author} {\bibinfo {author} {\bibfnamefont {O.}~\bibnamefont
  {Andreussi}}, \bibinfo {author} {\bibfnamefont {I.}~\bibnamefont {Dabo}}, \
  and\ \bibinfo {author} {\bibfnamefont {N.}~\bibnamefont {Marzari}},\
  }\href@noop {} {\bibfield  {journal} {\bibinfo  {journal} {J. Chem. Phys}\
  }\textbf {\bibinfo {volume} {136}},\ \bibinfo {pages} {064102} (\bibinfo
  {year} {2012})}\BibitemShut {NoStop}\bibitem [{\citenamefont {Dupont}, \citenamefont {Andreussi},\ and\
  \citenamefont {Marzari}(2013)}]{PCM-SCCS-charged}  \BibitemOpen
  \bibfield  {author} {\bibinfo {author} {\bibfnamefont {C.}~\bibnamefont
  {Dupont}}, \bibinfo {author} {\bibfnamefont {O.}~\bibnamefont {Andreussi}}, \
  and\ \bibinfo {author} {\bibfnamefont {N.}~\bibnamefont {Marzari}},\
  }\href@noop {} {\bibfield  {journal} {\bibinfo  {journal} {J. Chem. Phys}\
  }\textbf {\bibinfo {volume} {139}},\ \bibinfo {pages} {214110} (\bibinfo
  {year} {2013})}\BibitemShut {NoStop}\bibitem [{\citenamefont {Petrosyan}\ \emph {et~al.}(2007)\citenamefont
  {Petrosyan}, \citenamefont {Briere}, \citenamefont {Roundy},\ and\
  \citenamefont {Arias}}]{JDFT}  \BibitemOpen
  \bibfield  {author} {\bibinfo {author} {\bibfnamefont {S.~A.}\ \bibnamefont
  {Petrosyan}}, \bibinfo {author} {\bibfnamefont {J.-F.}\ \bibnamefont
  {Briere}}, \bibinfo {author} {\bibfnamefont {D.}~\bibnamefont {Roundy}}, \
  and\ \bibinfo {author} {\bibfnamefont {T.~A.}\ \bibnamefont {Arias}},\
  }\href@noop {} {\bibfield  {journal} {\bibinfo  {journal} {Phys. Rev. B}\
  }\textbf {\bibinfo {volume} {75}},\ \bibinfo {pages} {205105} (\bibinfo
  {year} {2007})}\BibitemShut {NoStop}\bibitem [{\citenamefont {Letchworth-Weaver}\ and\ \citenamefont
  {Arias}(2012)}]{PCM-Kendra}  \BibitemOpen
  \bibfield  {author} {\bibinfo {author} {\bibfnamefont {K.}~\bibnamefont
  {Letchworth-Weaver}}\ and\ \bibinfo {author} {\bibfnamefont {T.~A.}\
  \bibnamefont {Arias}},\ }\href@noop {} {\bibfield  {journal} {\bibinfo
  {journal} {Phys. Rev. B}\ }\textbf {\bibinfo {volume} {86}},\ \bibinfo
  {pages} {075140} (\bibinfo {year} {2012})}\BibitemShut {NoStop}\bibitem [{\citenamefont {Gunceler}\ \emph {et~al.}(2013)\citenamefont
  {Gunceler}, \citenamefont {Letchworth-Weaver}, \citenamefont {Sundararaman},
  \citenamefont {Schwarz},\ and\ \citenamefont {Arias}}]{NonlinearPCM}  \BibitemOpen
  \bibfield  {author} {\bibinfo {author} {\bibfnamefont {D.}~\bibnamefont
  {Gunceler}}, \bibinfo {author} {\bibfnamefont {K.}~\bibnamefont
  {Letchworth-Weaver}}, \bibinfo {author} {\bibfnamefont {R.}~\bibnamefont
  {Sundararaman}}, \bibinfo {author} {\bibfnamefont {K.}~\bibnamefont
  {Schwarz}}, \ and\ \bibinfo {author} {\bibfnamefont {T.}~\bibnamefont
  {Arias}},\ }\href@noop {} {\bibfield  {journal} {\bibinfo  {journal}
  {Modelling Simul. Mater. Sci. Eng.}\ }\textbf {\bibinfo {volume} {21}},\
  \bibinfo {pages} {074005} (\bibinfo {year} {2013})}\BibitemShut {NoStop}\bibitem [{\citenamefont {Gunceler}\ and\ \citenamefont
  {Arias}()}]{PCM-correlateTau}  \BibitemOpen
  \bibfield  {author} {\bibinfo {author} {\bibfnamefont {D.}~\bibnamefont
  {Gunceler}}\ and\ \bibinfo {author} {\bibfnamefont {T.~A.}\ \bibnamefont
  {Arias}},\ }\href@noop {} {\enquote {\bibinfo {title} {Universal iso-density
  polarizable continuum model for molecular solvents},}\ }\bibinfo {note}
  {Preprint arXiv:1403.6465v2}\BibitemShut {NoStop}\bibitem [{\citenamefont {Grimme}(2006)}]{Dispersion-Grimme}  \BibitemOpen
  \bibfield  {author} {\bibinfo {author} {\bibfnamefont {S.}~\bibnamefont
  {Grimme}},\ }\href@noop {} {\bibfield  {journal} {\bibinfo  {journal} {J.
  Comput. Chem}\ }\textbf {\bibinfo {volume} {27}},\ \bibinfo {pages} {1787}
  (\bibinfo {year} {2006})}\BibitemShut {NoStop}\bibitem [{\citenamefont {Ben-Amotz}\ and\ \citenamefont
  {Willis}(1993)}]{RvdwFluids}  \BibitemOpen
  \bibfield  {author} {\bibinfo {author} {\bibfnamefont {D.}~\bibnamefont
  {Ben-Amotz}}\ and\ \bibinfo {author} {\bibfnamefont {K.~G.}\ \bibnamefont
  {Willis}},\ }\href@noop {} {\bibfield  {journal} {\bibinfo  {journal} {J.
  Phys. Chem.}\ }\textbf {\bibinfo {volume} {97}},\ \bibinfo {pages} {7736}
  (\bibinfo {year} {1993})}\BibitemShut {NoStop}\bibitem [{\citenamefont {Sundararaman}, \citenamefont {Letchworth-Weaver},\
  and\ \citenamefont {Arias}(2012{\natexlab{a}})}]{JDFTx}  \BibitemOpen
  \bibfield  {author} {\bibinfo {author} {\bibfnamefont {R.}~\bibnamefont
  {Sundararaman}}, \bibinfo {author} {\bibfnamefont {K.}~\bibnamefont
  {Letchworth-Weaver}}, \ and\ \bibinfo {author} {\bibfnamefont {T.~A.}\
  \bibnamefont {Arias}},\ }\href@noop {} {\enquote {\bibinfo {title}
  {{JDFTx}},}\ }\bibinfo {howpublished} {\url{http://jdftx.sourceforge.net}}
  (\bibinfo {year} {2012}{\natexlab{a}})\BibitemShut {NoStop}\bibitem [{\citenamefont {Mantina}\ \emph {et~al.}(2009)\citenamefont
  {Mantina}, \citenamefont {Chamberlin}, \citenamefont {Valero}, \citenamefont
  {Cramer},\ and\ \citenamefont {Truhlar}}]{RvdwAtoms}  \BibitemOpen
  \bibfield  {author} {\bibinfo {author} {\bibfnamefont {M.}~\bibnamefont
  {Mantina}}, \bibinfo {author} {\bibfnamefont {A.~C.}\ \bibnamefont
  {Chamberlin}}, \bibinfo {author} {\bibfnamefont {R.}~\bibnamefont {Valero}},
  \bibinfo {author} {\bibfnamefont {C.~J.}\ \bibnamefont {Cramer}}, \ and\
  \bibinfo {author} {\bibfnamefont {D.~G.}\ \bibnamefont {Truhlar}},\
  }\href@noop {} {\bibfield  {journal} {\bibinfo  {journal} {J. Phys. Chem. A}\
  }\textbf {\bibinfo {volume} {113}},\ \bibinfo {pages} {5806} (\bibinfo {year}
  {2009})}\BibitemShut {NoStop}\bibitem [{\citenamefont {Pierotti}(1976)}]{SPT-Review}  \BibitemOpen
  \bibfield  {author} {\bibinfo {author} {\bibfnamefont {R.~A.}\ \bibnamefont
  {Pierotti}},\ }\href@noop {} {\bibfield  {journal} {\bibinfo  {journal}
  {Chem. Rev.}\ }\textbf {\bibinfo {volume} {76}},\ \bibinfo {pages} {717}
  (\bibinfo {year} {1976})}\BibitemShut {NoStop}\bibitem [{\citenamefont {Huang}, \citenamefont {Geissler},\ and\ \citenamefont
  {Chandler}(2001)}]{HardSphereSPCE}  \BibitemOpen
  \bibfield  {author} {\bibinfo {author} {\bibfnamefont {D.~M.}\ \bibnamefont
  {Huang}}, \bibinfo {author} {\bibfnamefont {P.~L.}\ \bibnamefont {Geissler}},
  \ and\ \bibinfo {author} {\bibfnamefont {D.}~\bibnamefont {Chandler}},\
  }\href@noop {} {\bibfield  {journal} {\bibinfo  {journal} {J. Phys. Chem. B}\
  }\textbf {\bibinfo {volume} {105}},\ \bibinfo {pages} {6704} (\bibinfo {year}
  {2001})}\BibitemShut {NoStop}\bibitem [{\citenamefont {Colominas}\ \emph {et~al.}(1999)\citenamefont
  {Colominas}, \citenamefont {Luque}, \citenamefont {Teixido},\ and\
  \citenamefont {Orozco}}]{Cavitation-Compare}  \BibitemOpen
  \bibfield  {author} {\bibinfo {author} {\bibfnamefont {C.}~\bibnamefont
  {Colominas}}, \bibinfo {author} {\bibfnamefont {F.~J.}\ \bibnamefont
  {Luque}}, \bibinfo {author} {\bibfnamefont {J.}~\bibnamefont {Teixido}}, \
  and\ \bibinfo {author} {\bibfnamefont {M.}~\bibnamefont {Orozco}},\
  }\href@noop {} {\bibfield  {journal} {\bibinfo  {journal} {Chem. Phys.}\
  }\textbf {\bibinfo {volume} {240}},\ \bibinfo {pages} {253} (\bibinfo {year}
  {1999})}\BibitemShut {NoStop}\bibitem [{\citenamefont {Sundararaman}, \citenamefont {Letchworth-Weaver},\
  and\ \citenamefont {Arias}(2012{\natexlab{b}})}]{BondedTrimer}  \BibitemOpen
  \bibfield  {author} {\bibinfo {author} {\bibfnamefont {R.}~\bibnamefont
  {Sundararaman}}, \bibinfo {author} {\bibfnamefont {K.}~\bibnamefont
  {Letchworth-Weaver}}, \ and\ \bibinfo {author} {\bibfnamefont {T.~A.}\
  \bibnamefont {Arias}},\ }\href@noop {} {\bibfield  {journal} {\bibinfo
  {journal} {J . Chem. Phys.}\ }\textbf {\bibinfo {volume} {137}},\ \bibinfo
  {pages} {044107} (\bibinfo {year} {2012}{\natexlab{b}})}\BibitemShut
  {NoStop}\bibitem [{\citenamefont {Sundararaman}\ and\ \citenamefont
  {Arias}(2014)}]{RigidCDFT}  \BibitemOpen
  \bibfield  {author} {\bibinfo {author} {\bibfnamefont {R.}~\bibnamefont
  {Sundararaman}}\ and\ \bibinfo {author} {\bibfnamefont {T.}~\bibnamefont
  {Arias}},\ }\href@noop {} {\bibfield  {journal} {\bibinfo  {journal} {Comp.
  Phys. Comm.}\ }\textbf {\bibinfo {volume} {185}},\ \bibinfo {pages} {818}
  (\bibinfo {year} {2014})}\BibitemShut {NoStop}\bibitem [{\citenamefont {Sundararaman}, \citenamefont {Letchworth-Weaver},\
  and\ \citenamefont {Arias}(2014)}]{PolarizableCDFT}  \BibitemOpen
  \bibfield  {author} {\bibinfo {author} {\bibfnamefont {R.}~\bibnamefont
  {Sundararaman}}, \bibinfo {author} {\bibfnamefont {K.}~\bibnamefont
  {Letchworth-Weaver}}, \ and\ \bibinfo {author} {\bibfnamefont {T.~A.}\
  \bibnamefont {Arias}},\ }\href@noop {} {\bibfield  {journal} {\bibinfo
  {journal} {J . Chem. Phys.}\ }\textbf {\bibinfo {volume} {140}},\ \bibinfo
  {pages} {144504} (\bibinfo {year} {2014})}\BibitemShut {NoStop}\bibitem [{\citenamefont {Langlet}\ \emph {et~al.}(1988)\citenamefont
  {Langlet}, \citenamefont {Claverie}, \citenamefont {Caillet},\ and\
  \citenamefont {Pullman}}]{Cavitation-CSPT}  \BibitemOpen
  \bibfield  {author} {\bibinfo {author} {\bibfnamefont {J.}~\bibnamefont
  {Langlet}}, \bibinfo {author} {\bibfnamefont {P.}~\bibnamefont {Claverie}},
  \bibinfo {author} {\bibfnamefont {J.}~\bibnamefont {Caillet}}, \ and\
  \bibinfo {author} {\bibfnamefont {A.}~\bibnamefont {Pullman}},\ }\href@noop
  {} {\bibfield  {journal} {\bibinfo  {journal} {J. Phys. Chem.}\ }\textbf
  {\bibinfo {volume} {92}},\ \bibinfo {pages} {1617} (\bibinfo {year}
  {1988})}\BibitemShut {NoStop}\bibitem [{\citenamefont {Amovilli}\ and\ \citenamefont
  {Mennucci}(1997)}]{RepulsionDispersionModel}  \BibitemOpen
  \bibfield  {author} {\bibinfo {author} {\bibfnamefont {C.}~\bibnamefont
  {Amovilli}}\ and\ \bibinfo {author} {\bibfnamefont {B.}~\bibnamefont
  {Mennucci}},\ }\href@noop {} {\bibfield  {journal} {\bibinfo  {journal} {J.
  Phys. Chem. B}\ }\textbf {\bibinfo {volume} {101}},\ \bibinfo {pages} {1051}
  (\bibinfo {year} {1997})}\BibitemShut {NoStop}\bibitem [{\citenamefont {Tkatchenko}\ and\ \citenamefont
  {Scheffler}(2009)}]{Dispersion-TS}  \BibitemOpen
  \bibfield  {author} {\bibinfo {author} {\bibfnamefont {A.}~\bibnamefont
  {Tkatchenko}}\ and\ \bibinfo {author} {\bibfnamefont {M.}~\bibnamefont
  {Scheffler}},\ }\href@noop {} {\bibfield  {journal} {\bibinfo  {journal}
  {Phys. Rev. Lett.}\ }\textbf {\bibinfo {volume} {102}},\ \bibinfo {pages}
  {073005} (\bibinfo {year} {2009})}\BibitemShut {NoStop}\bibitem [{\citenamefont {Tannor}\ \emph {et~al.}(1994)\citenamefont {Tannor},
  \citenamefont {Marten}, \citenamefont {Murphy}, \citenamefont {Friesner},
  \citenamefont {Sitkoff}, \citenamefont {Nicholls}, \citenamefont {Ringnalda},
  \citenamefont {Goddard},\ and\ \citenamefont
  {Honig}}]{solvation-exp-compiled-1}  \BibitemOpen
  \bibfield  {author} {\bibinfo {author} {\bibfnamefont {D.~J.}\ \bibnamefont
  {Tannor}}, \bibinfo {author} {\bibfnamefont {B.}~\bibnamefont {Marten}},
  \bibinfo {author} {\bibfnamefont {R.}~\bibnamefont {Murphy}}, \bibinfo
  {author} {\bibfnamefont {R.~A.}\ \bibnamefont {Friesner}}, \bibinfo {author}
  {\bibfnamefont {D.}~\bibnamefont {Sitkoff}}, \bibinfo {author} {\bibfnamefont
  {A.}~\bibnamefont {Nicholls}}, \bibinfo {author} {\bibfnamefont
  {M.}~\bibnamefont {Ringnalda}}, \bibinfo {author} {\bibfnamefont {W.~A.}\
  \bibnamefont {Goddard}}, \ and\ \bibinfo {author} {\bibfnamefont
  {B.}~\bibnamefont {Honig}},\ }\href@noop {} {\bibfield  {journal} {\bibinfo
  {journal} {J. Am. Chem. Soc.}\ }\textbf {\bibinfo {volume} {116}},\ \bibinfo
  {pages} {11875} (\bibinfo {year} {1994})}\BibitemShut {NoStop}\bibitem [{\citenamefont {Marten}\ \emph {et~al.}(1996)\citenamefont {Marten},
  \citenamefont {Kim}, \citenamefont {Cortis}, \citenamefont {Friesner},
  \citenamefont {Murphy}, \citenamefont {Ringnalda}, \citenamefont {Sitkoff},\
  and\ \citenamefont {Honig}}]{solvation-exp-compiled-2}  \BibitemOpen
  \bibfield  {author} {\bibinfo {author} {\bibfnamefont {B.}~\bibnamefont
  {Marten}}, \bibinfo {author} {\bibfnamefont {K.}~\bibnamefont {Kim}},
  \bibinfo {author} {\bibfnamefont {C.}~\bibnamefont {Cortis}}, \bibinfo
  {author} {\bibfnamefont {R.~A.}\ \bibnamefont {Friesner}}, \bibinfo {author}
  {\bibfnamefont {R.~B.}\ \bibnamefont {Murphy}}, \bibinfo {author}
  {\bibfnamefont {M.~N.}\ \bibnamefont {Ringnalda}}, \bibinfo {author}
  {\bibfnamefont {D.}~\bibnamefont {Sitkoff}}, \ and\ \bibinfo {author}
  {\bibfnamefont {B.}~\bibnamefont {Honig}},\ }\href@noop {} {\bibfield
  {journal} {\bibinfo  {journal} {J. Phys. Chem.}\ }\textbf {\bibinfo {volume}
  {100}},\ \bibinfo {pages} {11775} (\bibinfo {year} {1996})}\BibitemShut
  {NoStop}\end{thebibliography}
\end{document}